\documentstyle[prd,aps,floats,psfig]{revtex}
\begin{document}
\preprint{KNGU-INFO-PH-7, BROWN-HET-1175, hep-ph/yymmddd}
\draft

\renewcommand{\topfraction}{0.99}
\renewcommand{\bottomfraction}{0.99}
\twocolumn[\hsize\textwidth\columnwidth\hsize\csname 
@twocolumnfalse\endcsname

\title{\Large Stabilization of Embedded Defects by Plasma Effects}

\author{M. Nagasawa$^1$ and R. Brandenberger$^{2,3}$}
\address{~\\$^1$Department of Information Science, Faculty of Science,
Kanagawa University, Kanagawa 259-1293, JAPAN;
~\\$^2$Physics Department, Brown University, Providence, RI 02912,
USA; ~\\$^3$Department of Physics and Astronomy, University of British
Columbia, Vancouver, BC V6T 1Z1, CANADA}
\date{\today} 
\maketitle

\begin{abstract}
In models in which some of the scalar field are charged and some
uncharged, interactions with a finite temperature plasma will lead to
corrections to the effective potential of the charged fields which may
stabilize embedded defects made up of uncharged fields. In these
models embedded defects, solutions of the field equations which are
unstable at zero temperature, may thus be stable in the early
Universe, and may then play an important role in cosmology. Two
prototypical examples are the pion string in the theory of strong
interactions, and the electroweak Z-string in the standard electroweak
theory.
\end{abstract}

]

\section{Introduction}

Consider a quantum field theory with an n-dimensional vacuum manifold
${\cal M}_n$. An illustrative example which will be made use of in
most of this Letter is a theory of four real scalar fields $\phi_i,
\,\, i = 1, .., 4$, with a zero temperature potential
\begin{equation} \label{toy}
V(\phi) \, = \, {1 \over 4} \lambda \bigl( \sum_{i=1}^4 \phi_i^2 -
\eta^2 \bigr)^2
\end{equation}
which has as its vacuum manifold the three-dimensional sphere
${\cal M}_3 \, = \, S^3 \, = \, \{ (\phi_i): \sum_{i=1}^4 \phi_i^2 =
\eta^2 \}$. 

By freezing out certain combinations of the original fields we obtain
a sub-manifold ${\cal M}_m, \, m < n,$ of the original vacuum manifold
${\cal M}_n$. In the above example we can set $\phi_3 = \phi_4 = 0$
and obtain the one-dimensional sub-manifold
${\cal M}_1 \, = \, S^1 \, = \, \{ (\phi_i): \sum_{i=1}^2 \phi_i^2 =
\eta^2 \}$

As is well known$^{\cite{Kibble}}$, the topology of the vacuum
manifold determines the types of topological defects which the theory
admits. If the topology of ${\cal M}_m$ is such that k-dimensional
defects are possible (in four-dimensional space-time the criterion is
$\Pi_{2 - k}({\cal M}_m) \, \neq \, {\bf 1}$,
where $\Pi_l$ is the l'th homotopy group), then it is possible to
construct configurations of the unconstrained fields which correspond
to k-dimensional defects. If these configurations satisfy the
equations of motion of the unconstrained theory, we have an embedded
defect$^{\cite{TB,T92,BTB}}$. In the above example, the theory of
$(\phi_1, \phi_2)$ and vacuum manifold ${\cal M}_1$ admits linear
defects called cosmic strings. These string solutions with $\phi_3 =
\phi_4 = 0$ satisfy the full set of equations of motion and are hence
embedded strings.

Embedded defects are not topologically stable. In general, they are
not dynamically stable, either$^{\cite{instab}}$. The defect
configurations can unwind by exciting the frozen field
combinations. In the above example, the embedded string can unwind by
escaping into the $\phi_3$ and $\phi_4$ field dimensions.

If they were stable during a certain period in the early Universe,
embedded defects could play an important role in cosmology. The first
important example arises in the Standard Model of strong interactions,
which below the confinement scale is described by a sigma model
analogous to our toy model of Eq. (\ref{toy}), with $\phi_1$ denoting
the $\sigma$ field and $\phi_i, i = 2,3,4$ the three pions. This
theory admits no stable defects. Since $\Pi_3({\cal M}_3) \neq {\bf 1}$
there are $\Pi_3$ textures$^{\cite{Davis}}$ (these are $k = -1$
dimensional defects in the above notation), but in the
absence of a stabilizing Skyrme term $\Pi_3$ textures are
unstable$^{\cite{Turok}}$. However, by setting the charged pion fields
$\phi_3 = \phi_4 = 0$, we can construct embedded strings (pion
strings$^{\cite{pion}}$). A pion string along the z axis is given by
the configuration
\begin{equation} \label{string}
\phi(r, \theta) \, = \, \eta \rho(r) e^{i \theta}
\end{equation}
where $r, \theta$ are the polar coordinates in the x-y plane and $\phi
= \phi_1 + i \phi_2$. The profile function $\rho(r)$ satisfies the
boundary conditions $\rho(r) \rightarrow 0$ for $r \rightarrow 0$ and
$\rho(r) \rightarrow 1$ for $r \rightarrow \infty$. In Ref. \cite{BZ}
it was shown that due to the anomalous coupling with electromagnetism,
pion strings may provide a mechanism for generating primordial
galactic magnetic fields.

A second application of stabilized embedded defects is to
baryogenesis. Since defects trap energy in the unbroken state and
since they are out-of-equilibrium field configurations, they can play
an important role in baryogenesis$^{\cite{BDH}}$. Defect-mediated
baryogenesis may be implemented at the GUT scale$^{\cite{BDH}}$, at
the electroweak scale$^{\cite{BD,BDPT}}$, and possibly even at the QCD
scale$^{\cite{BHZ}}$. Embedded defects can contribute to baryogenesis
in the same way$^{\cite{BD}}$ as topological defects.  Consider, for
example, the standard electroweak theory based on the gauge group
$SU(2) \times U(1)$ with a complex Higgs doublet $\Phi = (\phi_+,
\phi_0)$, where the subscript indicates the electric charge. The
electroweak Z string$^{\cite{T92,Nambu}}$ is a solution of the
equations of motion obtained by embedding the
Nielsen-Olesen$^{\cite{NO}}$ $U(1)$ string in the following way:
\begin{equation}
\Phi \, = \, \rho_{NO}(r) e^{i \theta} (0, 1) \, , \,\,\,\, Z_{\mu} \, = \,
A_{\mu, NO} \, ,
\end{equation}
where $\rho_{NO}(r)$ and $A_{\mu, NO}$ are the scalar profile function
and the gauge field, respectively, of the Nielsen-Olesen string, and
$Z_{\mu}$ is the gauge field associated with the (electrically
uncharged) Z boson. It can be shown$^{\cite{instab}}$ that for
realistic values of the weak mixing angle, the electroweak string is
unstable at zero temperature. If stable, electroweak strings could
mediate electroweak scale baryon number violating
processes$^{\cite{BD}}$.

A third application of embedded defects is their role in mediating
interactions between stable topological defects. In Ref. \cite{ABES}
it has been shown that in a $O(3)$ linear sigma model there exists an
attractive force between monopoles and embedded walls and that upon
contact the monopole charge spreads out on the wall. Thus, embedded
walls can mediate a long range force between monopoles (they sweep up
monopoles and antimonopoles) and thus alleviate the monopole problem,
as proposed in the context of topological walls in
Ref. \cite{DLV}. The collapse of stable embedded defects could also
possibly lead to signatures in the X-ray and cosmic ray backgrounds.

There are many possible embedded defects - embedded
monopoles$^{\cite{BN}}$, embedded strings like the abovementioned pion
and electroweak strings, and embedded walls$^{\cite{BT}}$. In general,
the criteria for the existence of embedded defects are more
complicated$^{\cite{BTB}}$ than in our simple toy model. However, for
the rest of this Letter we shall stick to the simple model.

The formation of non-topological defects was studied in \cite{NY} and
\cite{ABD}. As shown in \cite{NY}, the formation probability of
electroweak strings is negligibly small if finite temperature effects
are neglected. In contrast, if embedded defects were stable at
temperatures below the symmetry breaking phase transition, they would
be produced in comparable abundance to that of topological defects of
the same dimension, by the usual Kibble argument$^{\cite{Kibble}}$.

As has been pointed out, it is possible that embedded defects can be
stabilized by bound states$^{\cite{VW,HHVW}}$ or by external electric$^{\cite{Periv}}$ and magnetic
fields$^{\cite{GM}}$. These mechanisms, however, are not generic. In
this Letter we propose a new and rather generic stabilization
mechanism which works whenever the fields excited in the embedded
defects are uncharged and the non-excited fields are charged, and when
the system is in a finite temperature plasma made up of charged
fields. In our examples, the charge will be the usual electric
charge. The mechanism, however, is more general and applicable to any
kind of charge, provided that the plasma is made up of particles which
carry that charge.

The basic idea is as follows: interactions with the charged plasma
will generate corrections to the effective potential for the scalar
fields which lift the potential in the directions of the charged
fields. This reduces the vacuum manifold ${\cal M}_n$ of the zero
temperature theory to a lower dimensional sub-manifold ${\cal M}_m$,
thus providing a way to stabilize embedded defects of the full theory
which are topological defects from the point of view of ${\cal M}_m$.

Our effect is different from an ordinary finite temperature effect. We
assume that the scalar fields are not in thermal equilibrium with the
plasma, but that the gauge fields which carry the charge force are, in
contrast to the usual framework of finite temperature field theory
where it is assumed that all fields are in thermal equilibrium. It
has been shown$^{\cite{HHVW}}$ that ordinary finite temperature
effects (which lift the effective potential in all field directions)
cannot in general stabilize embedded defects.

\section{Analytical Considerations}

As a toy model for the analytical study of the stabilization of
embedded defects by plasma effects we consider the chiral limit of the
QCD linear sigma model, involving the sigma field $\sigma$ and the
three pions ${\vec \pi} = (\pi^0, \pi^1, \pi^2)$, given by the
Lagrangian
\begin{equation} \label{lag1}
{\cal L}_0 \, = \, {1 \over 2} \partial_{\mu} \sigma \partial^{\mu}
\sigma + {1 \over 2} \partial_{\mu} {\vec \pi} \partial^{\mu} {\vec
\pi} - {\lambda \over 4} (\sigma^2 + {\vec \pi}^2 - v^2)^2 \, ,
\end{equation}
where $v^2$ is the ground state expectation value of $\sigma^2 + {\vec
\pi}^2$. In the following, we will denote the potential in
(\ref{lag1}) by $V_0$.

Two of the scalar fields, the $\sigma$ and $\pi_0$, are electrically
neutral, the other two are charged. Introducing the coupling to
electromagnetism, it is convenient to write the scalar field sector
${\cal L}$ of the resulting Lagrangian in terms of the complex scalar
fields
\begin{equation}
\pi^+ \, = {1 \over {\sqrt{2}}} (\pi^1 + i \pi^2) \, , \,\,\,
\pi^- \, = {1 \over {\sqrt{2}}} (\pi^1 - i \pi^2) \, .
\end{equation}
According to the minimal coupling prescription we obtain
\begin{equation} \label{lag2}
{\cal L} \, = \, {1 \over 2} \partial_{\mu} \sigma \partial^{\mu}
\sigma + {1 \over 2} \partial_{\mu} \pi^0 \partial^{\mu} \pi^0 +
D_{\mu}^+ \pi^+ D^{\mu -} \pi^- - V_0 \, ,
\end{equation}
where
\begin{equation} \label{partials}
D_{\mu}^+ \, = \, \partial_{\mu} + e A_{\mu} \, , \,\,\,\, D_{\mu}^-
\, = \, \partial_{\mu} - e A_{\mu} \, .
\end{equation}

If the gauge fields are part of the finite temperature plasma, we can
insert (\ref{partials}) into (\ref{lag2}) and do a Hartree-like
approximation by substituting
\begin{equation}
<A_{\mu}> \, = \, 0 \, , \,\,\,\, <A_{\mu}A^{\mu}> = \kappa T^2 \,
\end{equation}
where $T$ is the temperature and $\kappa$ is a numerical constant of
order unity, to extract from (\ref{lag2}) an effective potential for
the scalar fields of the form
\begin{equation} \label{poteff}
V_{T, eff}(\sigma, {\vec \pi}) \, = \, V_0 + \frac{1}{2}e^2\kappa T^2
((\pi^1)^2 + (\pi^2)^2) \, .
\end{equation}
At zero temperature, the vacuum manifold is $S^3$, but at finite
temperature it reduces to $S^1$. Thus, it is not unreasonable to
suspect that at high $T$ the embedded string (\ref{string}) is
stabilized.

To check the stability of the pion string (\ref{string}), we consider
the following variational ansatz for an unstable mode which obeys the
cylindrical symmetry of (\ref{string}) but escapes into the charged
field directions:
\begin{eqnarray} \label{var}
\phi = \sigma + i \pi^0 \, &=& \, v \rho(r) e^{i \theta} \, \,\,
\rho(r) \, = \, (1 - e^{- \mu r}) \\
\pi^1(r) \, &=& \, v \chi(r) \, , \nonumber
\end{eqnarray}
where $\mu$ is the width of the string (note that since the temperature only affects the charged fields and the embedded defects is made up of neutral
field, the width will be independent of $T$) and $\chi \ll 1$. We now
calculate the mass per unit length $I$ of the configuration
(\ref{var}) and compare it with the corresponding value $I_0$ for the
embedded string ($\chi = 0$). We obtain
\begin{equation}
I - I_0 \, = \, \pi v^2 \int_0^R dr r \chi^2 [e^2
\kappa T^2 - 2 \lambda v^2 e^{-\mu r}(1 - {1 \over 2} e^{-\mu r})] \, .
\end{equation}
The condition for stability is $I - I_0 > 0$ for all $\chi(r)$. As a
sufficient condition we get
\begin{equation} \label{bound}
T \, > \, 2 \lambda^{1/2} \kappa^{-1/2} e^{-1} v = T_D \, .
\end{equation}
If this temperature $T_D$ is smaller than the temperature $T_c$ of the
phase transition, then there will be a period $T \in [T_D, T_c]$
during which embedded defects are stable. The critical temperature
$T_c$ is obtained from the finite temperature effective
potential$^{\cite{FT}}$. Up to factors of order unity we obtain in our
toy model $T_c \simeq v$. Hence, provided that $\lambda e^{-2} \ll 1$
there is a period in the early Universe during which embedded defects
are stable.

So far, however, we have only studied the stability towards
perturbations maintaining the cylindrical symmetry. It is also
important to analyze the stability towards fragmentation. To do this,
we can consider the variational ansatz (\ref{var}) with a z-dependence
of $\rho$ and $\chi$. The only way this changes the previous analysis
is by adding to $I$ extra positive definite terms coming from spatial
gradient energies from z-derivatives. Hence, it is more difficult to
create local z-dependent perturbations than z-independent ones. Thus,
an embedded string with $T < T_D$ is also stable towards
fragmentation.

\section{Numerical Results}

We have simulated the formation of embedded defects in the presence of
a charged plasma using a code based on the one employed in
\cite{NSY}. Four scalar fields $\phi$ are evolved numerically on a
three-dimensional lattice by means of the equations of motion derived
from (\ref{lag1}) with the potential $V_0$ replaced by the effective
potential (\ref{poteff}). All dimensional quantities were rescaled by
appropriate powers of $v$ to make them dimensionless. For most
simulations, a box size of $50^3$ was used (we checked that the basic
results were insensitive to the box size by executing $100^3$ and
$200^3$ box simulations.). The spatial resolution was
$\Delta x = v^{-1}$, and the time steps were chosen as $\Delta t = {1
\over 100} \Delta x$.  We used two different sets of initial
conditions. The results shown below are for {\it true vacuum} initial
conditions, choosing the fields to be randomly distributed (on a
length scale of $\Delta x$) over the entire vacuum manifold ${\cal
M}_n = S^3$:
$\vert \phi(x) \vert \, = \, v \, , \,\,\, {\dot \phi}(x) \, = \, 0$.
We also performed simulations with {\it false vacuum} initial
conditions determined by
$\phi(x) \, = \, 0 \, , \,\,\, \vert {\dot \phi}(x) \vert \, = \, T^2 \, ,$
with random orientations of the direction of ${\dot \phi}$ in the
field tangent space.

\begin{figure}
\begin{center}
\mbox{\psfig{figure=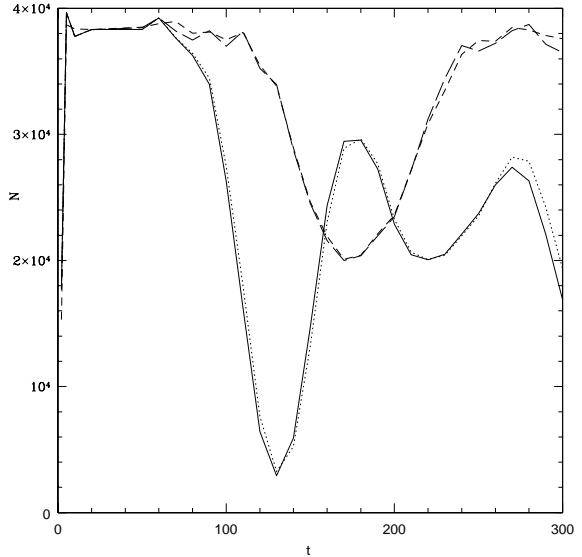,height=8cm}}
\end{center}
\caption{Time evolution of the number $N$ of points where $\vert\phi\vert
<0.5v$. The simulations start at $t=1$ and the result
is the averaged value over 20  simulations for different initial
phase configuration in a $50^3$ box. The solid line corresponds to
$T=10^{-3}v$, the dotted one to $T=10^{-2}v$, the long-dashed
one to $T=10^{-1}v$, and the short-dashed one to $T=v$.}
\end{figure}

The presence of embedded defects implies the existence of points in
space with $\vert \phi(x) \vert < \alpha v$, where $\alpha < 1$ is an
arbitrary constant which determines what is meant by the core radius
of the defect (and chosen to be $\alpha = 1/2$ in Figs. 1 and 2). For
true vacuum initial conditions, we expect that the number $N$ of grid
points which satisfy the above criterion is initially independent of
the temperature $T$ (and simply indicative of the random fluctuations
set up by the initial gradient energy). As a function of time, we
expect the number of points to decrease if defects are unstable, but
to converge to a constant value if defects are stable. Thus, we expect
an abrupt increase in $N$ at later times at the stabilization
temperature $T_D$. We observed this in our simulations. Fig. 1 shows
the results for the case $\kappa = e = 1$ and $\lambda = 10^{-2}$. The
numerically determined value of $T_D$ lies between $10^{-2}v$ and
$10^{-1}v$, in good agreement with the analytical upper bound of
(\ref{bound}). The fact that there are points with $\vert \phi(x) \vert
< \alpha v$ for all temperatures is due to the large energy in the
system due to the initial gradient energy which leads to
fluctuations. In an expanding Universe, these fluctuations would be
damped. However, expansion is not included in our code.

We can introduce a damping term in order to reduce the effects of fluctuations.
In Fig. 2 we compare the three-dimensional distribution of
points with $\vert \phi(x) \vert < \alpha v$ in a simulation
with $T = v$ (stable defects expected) and with $T = 10^{-3}v$
(no stable defects expected). In the first case, although the zero
point constraint is more stringent than the second one due to
the smaller $alpha$, strings of grid
points with $\vert \phi(x) \vert < \alpha v$ are apparent, whereas
such strings are absent in the second case.

\begin{figure}
\begin{center}
\mbox{\psfig{figure=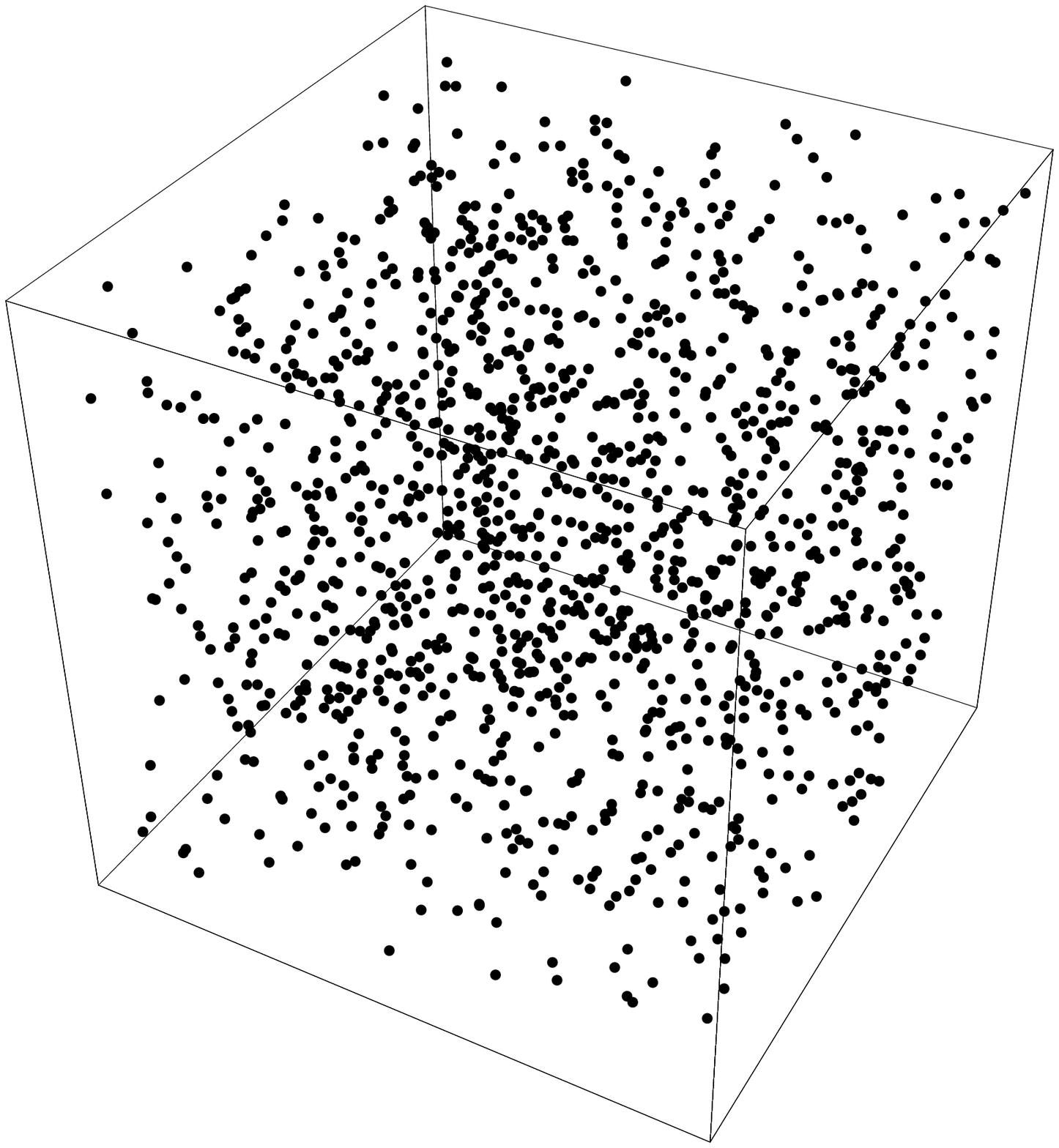,height=5cm}}
\quad\quad \mbox{\psfig{figure=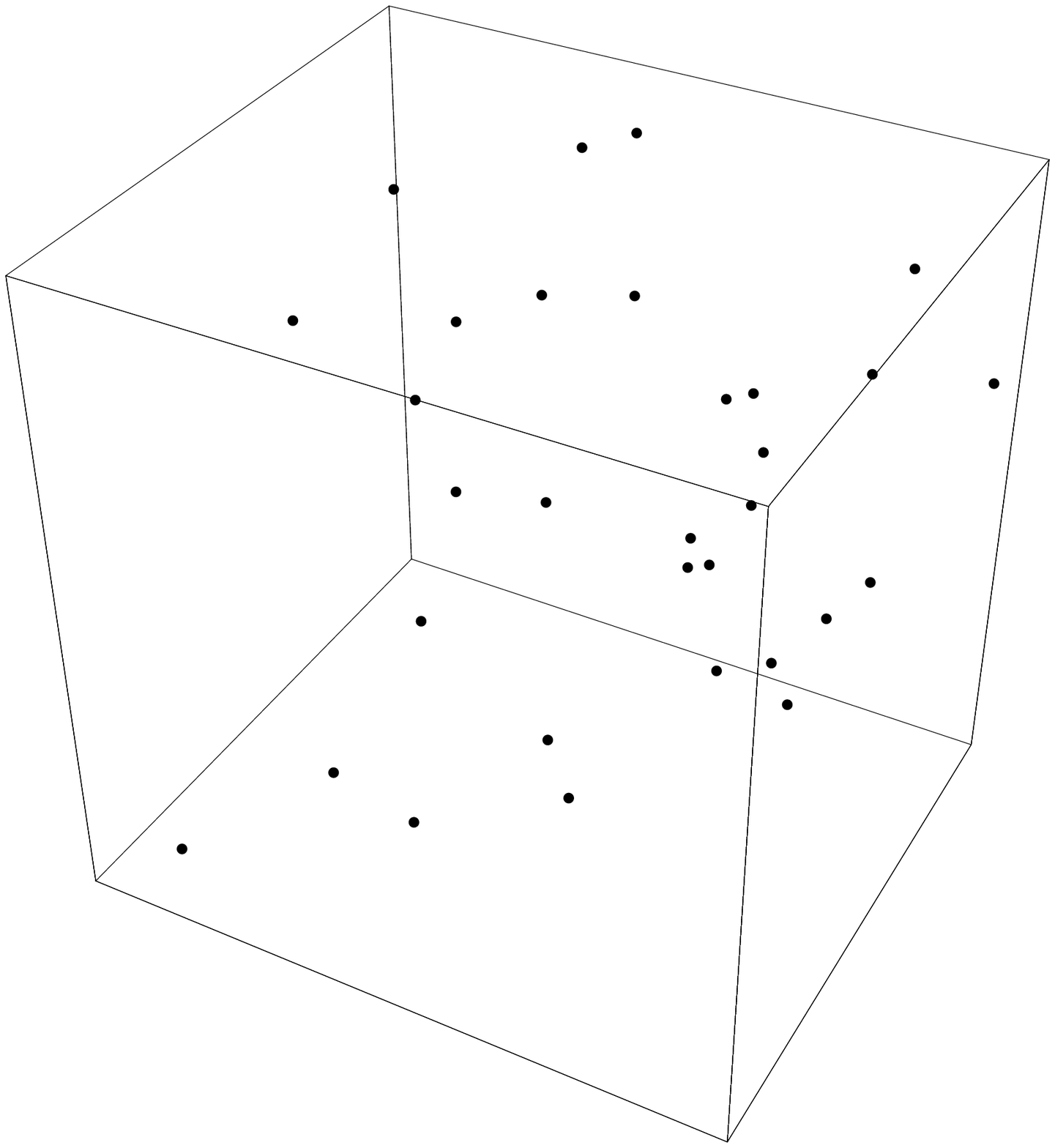,height=5cm}}
\end{center}
\caption{The three-dimensional distribution of cells where
$\vert \phi \vert < \alpha v$ at $t=80$ for simulations including
the damping term. The left graph is for $T=v$, $\alpha=0.1$
and the right one for $T=10^{-3}v$, $\alpha=0.15$.}
\end{figure}

\section{Discussion}

We have discussed the stabilization of embedded defects made up of
neutral fields in the presence of a charged plasma. We illustrated the
effect in the context of pion strings which can be stabilized by the
electromagnetic plasma. However, the same mechanism applies to other charges,
for example to the color-charged plasma above the QCD confinement scale which has the potential of stabilizing color-neutral embedded defects.

The stabilization of neutral embedded defects is due to finite temperature plasma effects which add (to leading order) quadratic corrections to the effective potential for the charged fields. No external charged fields (as in the stabilization mechanism of \cite{Periv} in which embedded defects which couple to the external charge are stabilized by an induced angular momentum) are required. 

For our mechanism to operate it is important that the fields which make up the embedded defects are no longer in thermal equilibrium whereas the gauge fields which carry the charges are still thermally excited. This condition is easier to realize for global defects such as pion strings than for local defects such as electroweak strings in which the gauge fields themselves play an important role.

Important questions for future study are the formation probability and length distribution of stabilized embedded strings. Unlike topologically stable strings, stabilized embedded strings may have ends (in which case they are
expected to be short-lived). Answers to these questions will be crucial in order to be able to discuss the cosmological consequences of the scenario. Work on this topic is in progress. 
    
\centerline{Acknowledgments}

MN acknowledges the kind hospitality of Brown University and University of
British Columbia (UBC) where most of this work was done. R.B. thanks W. Unruh for hospitality at UBC. We are grateful to T. Vachaspati for comments on a draft of the paper.
The work of R.B. is supported in part by the U.S. Department of Energy under
Contract DE-FG02-91ER40688, TASK A.


\begin{thebibliography} {99}
  
\bibitem{Kibble} T.W.B. Kibble, {\it J. Phys.} {\bf A9}, 1387 (1976).

\bibitem{TB} T. Vachaspati and M. Barriola, {\it Phys. Rev. Lett.} {\bf 69}, 1867 (1992).

\bibitem{T92} T. Vachaspati, {\it Phys. Rev. Lett.} {\bf 68}, 1977 (1992);
{\it Nucl. Phys.} {\bf B397}, 648 (1993).

\bibitem{BTB} M. Barriola, T. Vachaspati and M. Bucher, {\it Phys. Rev.} {\bf D50}, 2819 (1994).

\bibitem{instab} M. James, L. Perivolaropoulos and T. Vachaspati, {\it Phys. Rev.}
{\bf D46}, 5232 (1992); {\it Nucl. Phys.} {\bf B395}, 534 (1993);
M. Goodband and M. Hindmarsh, {\it Phys. Lett.} {\bf B363}, 58 (1995).

\bibitem{Davis} R. Davis, {\it Phys. Rev.} {\bf D35}, 3705 (1987).

\bibitem{Turok} N. Turok, {\it Phys. Rev. Lett.} {\bf 63}, 2625 (1990).

\bibitem{pion} X. Zhang, T. Huang and R. Brandenberger, {\it Phys. Rev.} {\bf D58}, 027702 (1998), hep-ph/9711452. 

\bibitem{BZ} R. Brandenberger and X. Zhang, {\it Phys. Rev. D}, in press (1999),
hep-ph/9808306.

\bibitem{BDH} R. Brandenberger, A.-C. Davis and M. Hindmarsh, {\it Phys. Lett.}
{\bf B263}, 239 (1991).

\bibitem{BD} R. Brandenberger and A.-C. Davis, {\it Phys. Lett.} {\bf B308}, 79 (1993).

\bibitem{BDPT} R. Brandenberger, A.-C. Davis and M. Trodden, {\it Phys. Lett.} {\bf B335}, 123 (1994); R. Brandenberger, A.-C. Davis, T. Prokopec and M. Trodden, {\it Phys. Rev.} {\bf D53}, 4257 (1996).

\bibitem{BHZ} R. Brandenberger, I. Halperin and A. Zhitnitsky, hep-ph/9808471. 

\bibitem{NO} H. Nielsen and P. Olesen, {\it Nucl. Phys.} {\bf B61}, 45 (1973).

\bibitem{Nambu} Y. Nambu, {\it Nucl. Phys.} {\bf B130}, 505 (1977);
K. Huang and R. Tipton, {\it Phys. Rev.} {\bf D23}, 3050 (1981);
N. Manton, {\it Phys. Rev.} {\bf D28}, 2019 (1983).

\bibitem{ABES} S. Alexander, R. Brandenberger, R. Easther and A. Sornborger, hep-ph/9903254.

\bibitem{DLV} G. Dvali, H. Liu and T. Vachaspati, {\it Phys. Rev. Lett.} {\bf 80}, 2281 (1998).

\bibitem{BN} R. Brandt and F. Neri, {\it Nucl. Phy.} {\bf B161}, 253 (1979).

\bibitem{BT} C. Bachas and T. Tomaras, {\it Phys. Rev. Lett.} {\bf 76}, 356 (1996).

\bibitem{NY} M. Nagasawa and J. Yokoyama, {\it Phys. Rev. Lett.} {\bf 77}, 2166 (1996).

\bibitem{ABD} A. Achucarro, J. Borrill and A. Liddle, hep-ph/9802306.

\bibitem{VW} T. Vachaspati and R. Watkins, {\it Phys. Lett.} {\bf B318}, 163 (1993).

\bibitem{HHVW} R. Holman, S. Hsu, T. Vachaspati and R. Watkins, {\it Phys. Rev.}
{\bf D46}, 5352 (1992).

\bibitem{Periv} L. Perivolaropoulos, {\it Phys. Rev.} {\bf D50}, 962 (1994).

\bibitem{GM} J. Garriga and X. Montes, {\it Phys. Rev. Lett.} {\bf 75}, 2268 (1995).

\bibitem{FT} L. Dolan and R. Jackiw, {\it Phys. Rev.} {\bf D9}, 3320 (1974);
for a review see e.g. R. Brandenberger, {\it Rev. Mod. Phys.} {\bf 57}, 1 (1985).

\bibitem{NSY} M. Nagasawa, K. Sato and J. Yokoyama, {\it Publ. Astro. Soc. of Japan}
{\bf 45}, 755 (1993).

\end{thebibliography}
\end{document}